\documentclass[12pt,preprint]{aastex}
\begin{document}

\shortauthors{Stanek}

\shorttitle{Citations vs.~number of authors}

\title{Are astronomical papers with more authors cited more?}

\author{K. Z. Stanek\altaffilmark{1}}

\altaffiltext{1}{\small Department of Astronomy, The Ohio State University, Columbus, OH 43210}

\email{kstanek@astronomy.ohio-state.edu}

\begin{abstract}

Following my previous study of paper length vs. number of citations in
astronomy (Stanek 2008), some colleagues expressed an interest in
knowing if any correlation exists between citations and the number of
authors on an astronomical paper.  At least naively, one would expect
papers with more authors to be cited more. I test this expectation
with the same sample of papers as analyzed in Stanek (2008), selecting
all ($\sim 30,000$) refereed papers from A\&A, AJ, ApJ and MNRAS
published between 2000 and 2004. These particular years were chosen so
that the papers analyzed would not be too ``fresh'', but number of
authors and length of each article could be obtained via ADS.

I find that indeed papers with more authors published in these four
major astronomy journals are on average cited more, but only weakly
so: roughly, the number of citations doubles with ten-fold increase in
the number of authors. While the median number of citations for a 2
author paper is 17, the median number of citations to a paper with 10
to 20 authors is 32. However, I find that most papers are written by a
small number of authors, with a mode of 2 authors and a median of 3
authors, and 92\% of all papers written have fewer than 10 authors.
Perhaps surprisingly, I also find that papers with more authors are
not longer than papers with fewer authors, in fact a median number of
8 to 10 pages per paper holds for any number of authors.  For the same
sample of papers, a median number of citations per paper grew from 15
in June 2008 (Stanek 2008) to 19 in November 2009. Unlike Stanek
(2008), I do not conclude with any career advice, semi-humorous or
otherwise.

\end{abstract}

\section{Introduction}

\addtocounter{footnote}{1}

There have been a number of publications analyzing citation patterns
in the astronomical literature, see my previous paper for more
discussion (Stanek 2008). In that paper I addressed the question of
whether longer astronomical papers are cited more (they are). Following
a number of suggestions I received after posting that
seminal\footnote{Yes, I am joking} study, here I address the question
of whether astronomical papers with more authors are also cited more.
That this would be the case is somewhat expected, for a number of
reasons, but the expected size of the effect varied quite widely among
my informally polled colleagues. I therefore decided to investigate
the citation vs.~number of authors correlation for astronomical
papers, and I found the results interesting enough to warrant this
posting. In Section 2 I describe the data, namely citation and number
of authors statistics for about 30,000 refereed astronomical papers
published between 2000 and 2004 in ApJ, A\&A, AJ and MNRAS. In Section
3 I discuss the results, and in Section 4 I conclude with some brief
remarks.

\section{Data}

I used the ADS Abstract Service\footnote{{\tt
http://adsabs.harvard.edu/abstract$\_$service.html}} to obtain the
citations, page length and number of authors for all papers published
between 2000 and 2004 (i.e., five complete years) in the four largest
astronomical journals, A\&A, AJ, ApJ and MNRAS.\footnote{These data
are available by request from the Author} I start in 2000 so that page
lengths will be reported for all papers and end with 2004 to allow
significant time for papers to mature (or not) and obtain citations.
The journals were chosen so there would be plenty of papers for
statistics, with the page length being about the same for each
journal.  There were 12,854 ApJ (including ApJL and ApJS) papers,
9,275 A\&A papers, 5,328 MNRAS papers and 2,564 AJ papers, for a total
of 30,021 papers. Originally, that number was somewhat larger, but I
have removed all one-page papers from the list, as these were mostly
errata or editorials, usually cited 0 times. I procured these data
using ADS in mid-November 2009. As this is the same sample of papers I
have used in my previous study (Stanek 2008), but previously with
citations accumulated until June 2008, I will also discuss briefly the
growth of citations for the whole sample between the two studies.

\begin{figure}[p]
\plotone{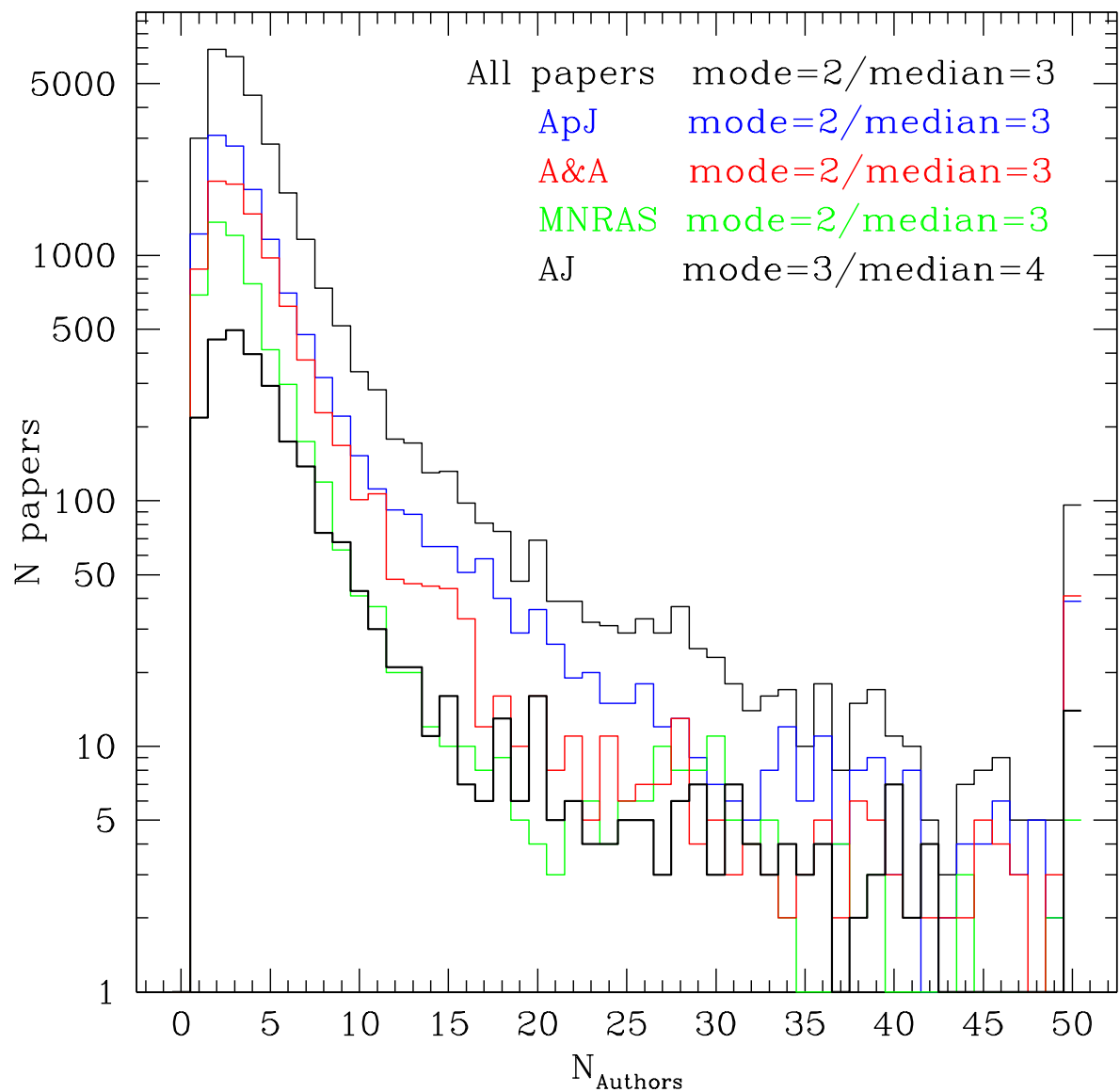}
\caption{Distribution of the number of authors for 30,021 A\&A, AJ,
ApJ, and MNRAS papers published between 2000--2004. There are about
100 papers with 50 or more authors, and there are about 3,000 papers
with one author.}
\label{fig1}
\end{figure}

\begin{figure}[p]
\plotone{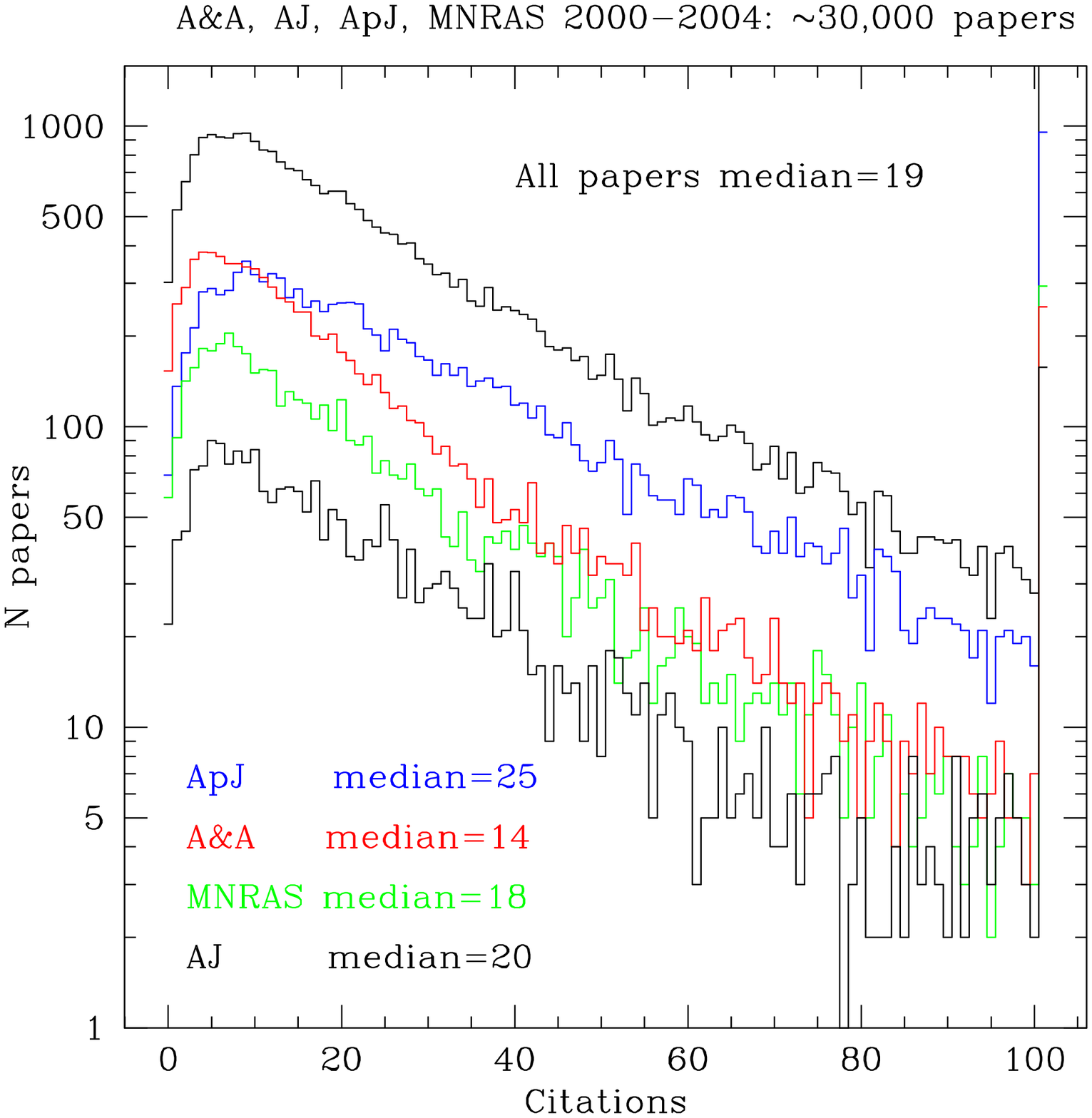}
\caption{Distribution of citations for 30,021 A\&A, AJ, ApJ and MNRAS
papers published between 2000--2004, shown for the entire sample and
for each journal separately. This is analogous figure to Fig.$\;$2 in
Stanek (2008), but with citations numbers updated for November 2009.
Median number of citations for the entire sample grew from 15 in June
2008 to 19 in November 2009. There are about 1,700 papers with 100 or
more citations and 10 papers with more than 1000 citations.}
\label{fig2}
\end{figure}

In Figures 1 and 2 I show the basic properties of our sample, such as
the distribution of number of authors (Fig.$\,$1) and the distribution
of citations (Fig.$\,$2). While the distributions presented in
Fig.$\,$1 are more or less the same, except for AJ having a larger
fraction of papers with more authors, the distributions presented in
Fig.$\,$2 are not identical in shape.

The median number of authors per paper for the entire sample of
$\sim30,000$ papers is 3, with a mode of 2 authors, and only $\sim100$
papers with 50 or more authors. In the distribution of citations,
$\sim2,3000$ papers in the sample have 3 or fewer citations, while
$\sim1,700$ papers have 100 or more citations, and only $10$ papers
have 1,000 or more citations.  Th median number of citations for the
entire sample grew from 15 in June 2008 (Stanek 2008) to 19 in
November 2009. I have not tried in any way to correct for or to remove
self-citations.

\section{Results} 

Having the number of authors and the number of citations, I can
produce a Citation-Authors Diagram (hereafter: CAD), which I present
in Fig.$\,$3. For display purposes, I have added $+1$ to the number of
citations for each paper (``an honorary citation''), to avoid the
$\log{citations}=-\infty$ problem.  Also, for display purposes, I add
a random variable uniformly distributed between 0 and 0.9 to both
$N_{Authors}$ and the number of citations, otherwise all papers with
the same number of authors and citations would produce only one dot in
the CAD.

For bins of journals articles with a given number of authors, the
median number of citations is shown by large open circles. As can be
seen from our CAD, these medians clearly increase with the number of
authors, and range from 12 for articles with 1 author to about 50 for
articles with 50 and more authors.  One obvious thing to notice is
that, as expected, for any given number of authors on a paper there is
a very wide range in the number of citations, as seen by the 10\% and
90\% ranges (filled blue dots). Somewhat surprisingly, barring any ADS
errors, there seem to exist papers with very many authors and very few
citations.

Since I have already shown (Stanek 2008) that longer papers are cited
more (again, in the median sense, and as always with a large
dispersion in number of citations), maybe what we are seeing in
Fig.$\;$3 is simply that having more authors also means writing more
pages? However, as can be seen in Fig.$\;$4, that is definitely not
the case, in fact the median number of pages does not seem to depend
at all on the number of authors and it is always from 8 to 10
pages. Additional authors clearly do not seem to be actually
contributing much in terms of additional paper length, but they are
clearly adding (logarithmically) to a paper's impact.  This additional
impact could be simply due to more self-citations, but that is only a
guess, which I have not investigated here.

Finally, in Fig$\;$5 I plot the cumulative distributions of the
fraction of the total citations vs. number of authors for each of the
four journals. The curves for ApJ, A\&A and MNRAS are rather similar,
while the cumulative curve for AJ is significantly different, meaning
papers with more authors contribute a significantly larger fraction of
total citations for AJ vs. the other journals. I will let the Reader
amuse herself/himself by speculating on or investigating the possible
reasons for this difference.

\begin{figure}[p]
\plotone{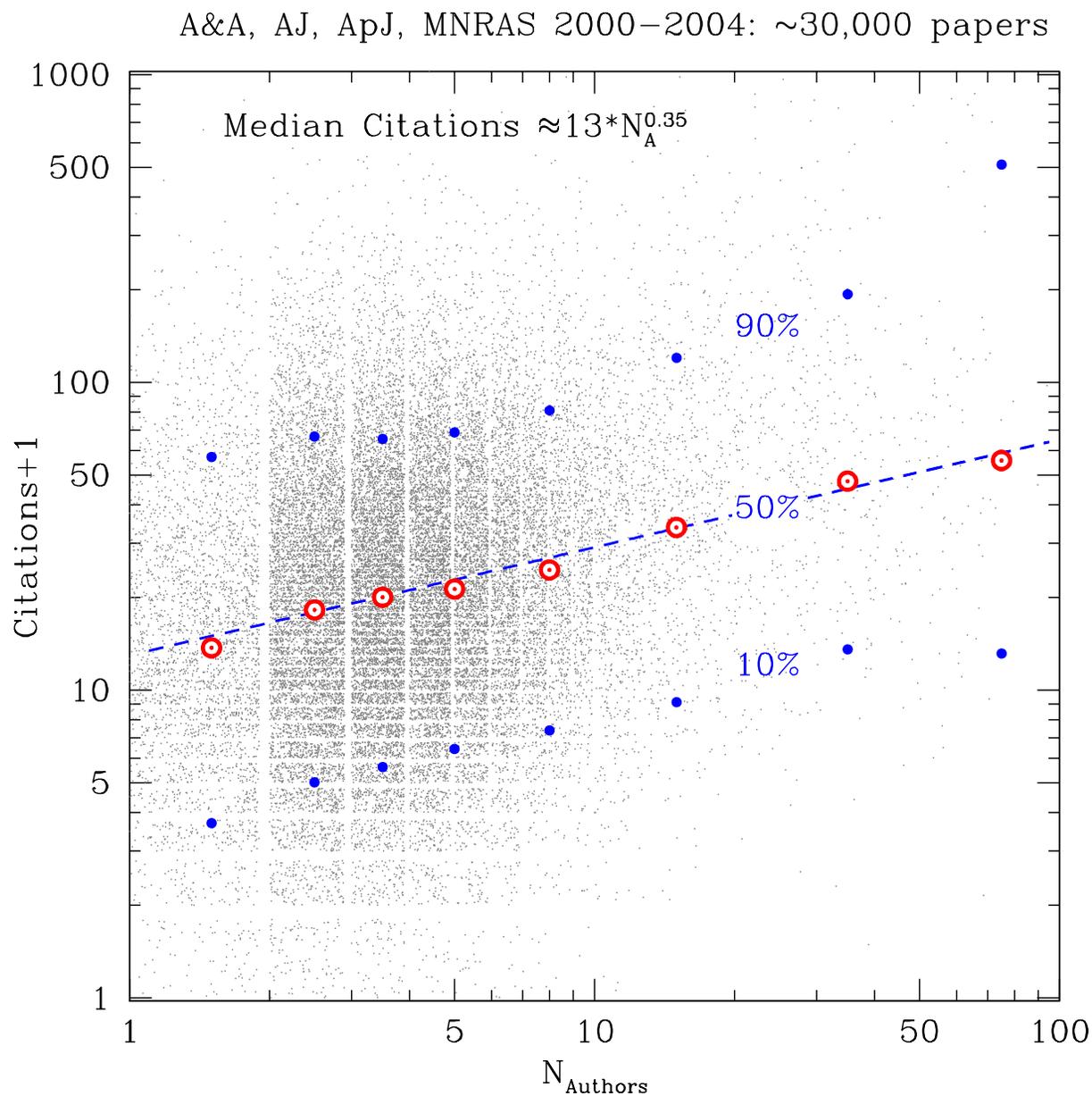}
\caption{Citations vs.~number of authors for our sample of 30,021
papers published in A\&A, AJ, ApJ, and MNRAS between 2000--2004. For
bins of articles with a given number of authors, the median number of
citations is shown with the large dotted circle, ranging from 12
citations for articles with $1$ author to $\sim50$ median citations
for articles with more than $50$ authors (there are only $\sim100$
papers in that last bin). Also shown are 10\% and 90\% ranges in each
bin. The ``best-fit'' line was fit by eye.}
\label{fig3}
\end{figure}

\begin{figure}[p]
\plotone{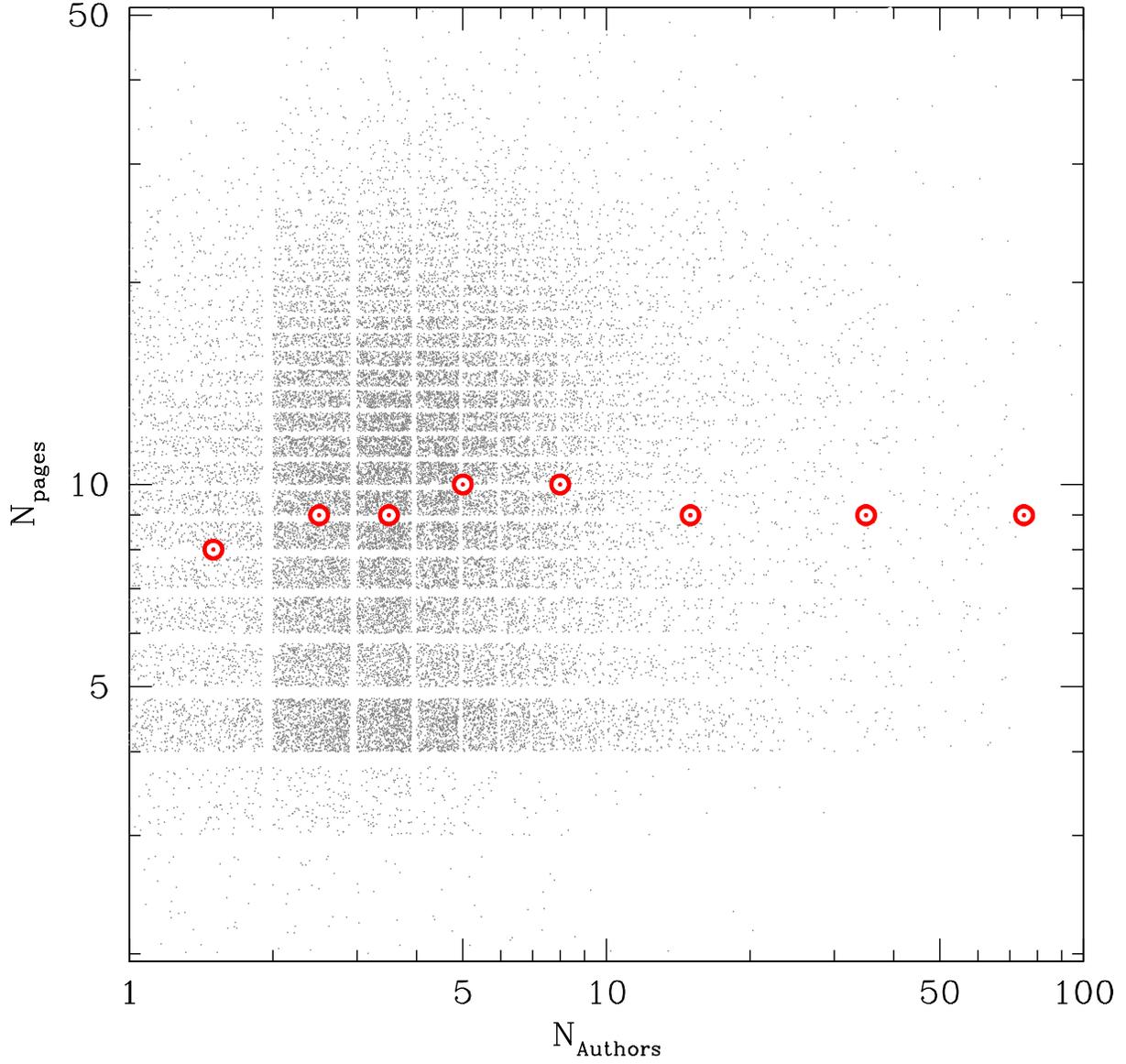}
\caption{Length of paper vs. number of authors on a paper. The median
number of pages in each bin is shown with the large dotted circle.}
\label{fig4}
\end{figure}

\begin{figure}[p]
\plotone{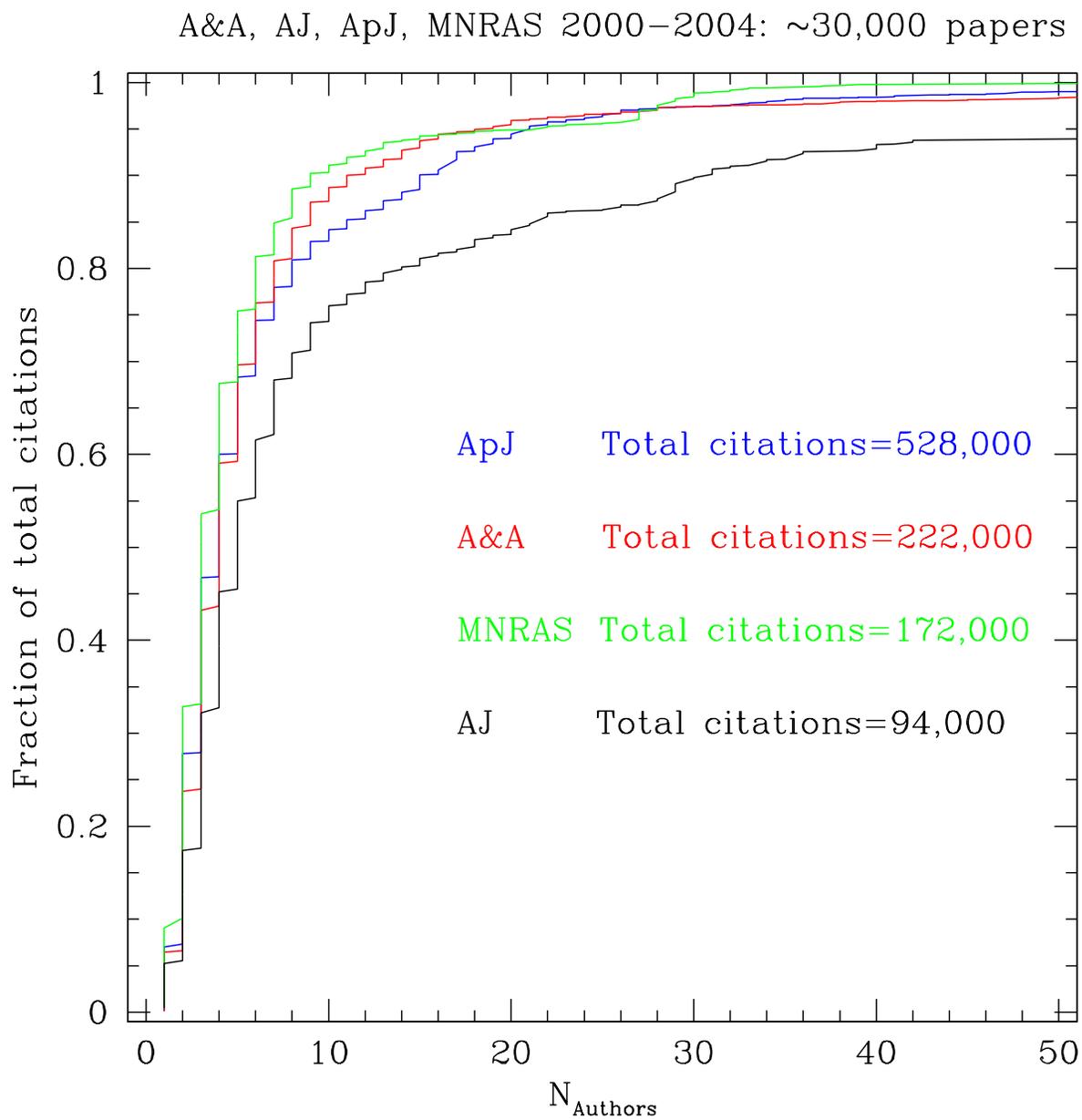}
\caption{Cumulative distributions of the total citations vs. number of
authors for each of the four journals.}
\label{fig5}
\end{figure}

\section{Brief Remarks}

In Stanek (2008) I concluded with some semi-humorous publishing
advice, directed especially towards first-year graduate students.  I
stand by that advice, also because of the mostly positive feedback I
have received after posting that study\footnote{Yes, it is a biased sample}:
``I enjoyed your well written article about citations and the length
of papers. Perhaps surprisingly, the longest of the $\sim$N00 papers
that I have written was often cited, but not much read. So perhaps one
should look for ways to try to measure the impact of a paper rather
than the number of times it is cited - clearly a difficult task.
Perhaps the truly great papers are the ones that still get cited after
many decades. It is a bit frightening to note that Zwicky's great
paper on the discovery of dark matter only started to get cited a
quarter of a century after it was written.''; ``I wanted to say that I
read your paper and really enjoyed it.  In particular, the last
section where you gave `semi-humorous' advice.  That was the best
advice I've heard in a long time!''; ``brilliant astro-ph submission!
I think this will be a mandatory read for any students that want to go
on to grad school.'; ``Thanks for your insightful and very funny paper
on astro-ph.  It made my morning.''; but also ``Cute paper on astro-ph
today, but don't you find this whole series of meta-papers to be
rather cynical? You call the advice to first-year graduate students
like myself `semi-humorous,' but the message seems to be telling us
how to game the system.''

I do not have enough new semi-humorous material to put in this paper,
but if you send me something funny and relevant, I might add you and
the advice to the paper, and perhaps together we will be cited more
(logarithmically speaking). It would perhaps be of some interest to
investigate how the effect found in this paper changes with time, and
more generally if the statistics of number of authors has evolved in
the last few decades. This might or might not be a subject of a future
study.

\acknowledgments

I thank Hsiao-Wen Chen and Chris Kochanek for suggesting this
particular line of inquiry. I also thank Chris Kochanek for locating
all these ``the''s and ``a''s that I have lost, misplaced or applied
randomly. I thank the people whose e-mails I am quoting (I will keep
them anonymous). The data and plots presented in this paper are real,
despite the (mostly intended) humorous style of writing. This paper
will not be submitted to any journal.

\end{document}